\documentclass[10pt]{article}

\usepackage[hidelinks]{hyperref}
\usepackage{amsmath,amsthm,amsfonts,amssymb,mathtools}
\usepackage{float}
\usepackage{graphicx}
\usepackage{epstopdf}
\usepackage{tcolorbox}

\usepackage{amsmath}
  \usepackage{paralist}
  \usepackage{subfig}
\usepackage{tikz-cd} 
\usepackage{mathrsfs}

\definecolor{dullmagenta}{RGB}{102,0,102}

\makeatletter 
\renewcommand\paragraph{\@startsection{paragraph}{4}{\z@}%
                                      {\parskip}
                                      {-1em}%
                                      {\normalfont\normalsize\bfseries}}
\makeatother

\makeatletter
\def\blfootnote{\gdef\@thefnmark{}\@footnotetext}
\makeatother

\parskip=\medskipamount
\makeatletter 
\def\@endtheorem{\endtrivlist}
\makeatother

\makeatletter
\def\th@plain{%
  \thm@notefont{}
  \itshape 
}
\def\th@definition{%
  \thm@notefont{}
  \normalfont 
}
\makeatother

\theoremstyle{plain}

\newtheorem{proposition}{Proposition}

\theoremstyle{definition}
\newtheorem{remark}{Remark}

\newtheorem{definition}{Definition}
\newtheorem*{theorem*}{Theorem}

\begin{document}
\date{}
\title{A note on Hybrid Routh reduction for time-dependent Lagrangian systems}

\author{L.\ J.\ Colombo\textsuperscript{a}, M.\ E.\ Eyrea Iraz\'u\textsuperscript{b} and  E.\ Garc\'{\i}a-Tora\~{n}o Andr\'{e}s\textsuperscript{c}\\[2mm]
{\small \textsuperscript{a} Instituto de Ciencias Matem\'aticas, Consejo Superior de Investigaciones Cient\'ificas} \\
{\small  Calle Nicol\'as Cabrera 13-15, Cantoblanco, 28049, Madrid, Spain}\\[2mm]
{\small \textsuperscript{b} Departamento de Matem\'atica, Universidad Nacional de La Plata} \\
{\small  Calle 1 y 115, La Plata 1900, Buenos Aires, Argentina}\\[2mm]
{\small \textsuperscript{c} Departamento de Matem\'atica,
Universidad Nacional del Sur}  \\
{\small  Av.\ Alem 1253, 8000 Bah\'ia Blanca, Argentina}} 

\maketitle
\blfootnote{L. Colombo was partially supported by I-Link Project (Ref: linkA20079) from CSIC, Ministerio de Economia, Industria y Competitividad (MINEICO, Spain) under grant MTM2016- 76702-P; ``Severo Ochoa Programme for Centres of Excellence'' in R$\&$D (SEV-2015-0554). The project that gave rise to these results received the support of a fellowship from ``La Caixa' Foundation (ID 100010434). M.E. Eyrea Iraz\'u was partially supported by CONICET Argentina.}

\begin{abstract}

This note discusses Routh reduction for hybrid time-dependent mechanical systems. We give general conditions on whether it is possible to reduce by symmetries a hybrid time-dependent Lagrangian system extending and unifying previous results for continuous-time systems. We illustrate the applicability of the method using the example of a billiard with moving walls.

\vspace{3mm}

\textbf{Keywords:} Symmetries, cyclic coordinates, hybrid mechanical systems, conserved quantities, cosymplectic reduction.

\vspace{3mm}

\textbf{2010 Mathematics Subject Classification:}  70S10,  37J15, 70H03.
\end{abstract}

\section{Introduction}

One of the first instances of symmetry reduction for mechanical systems can be found in the pioneering works of Routh in the second half of the 19th century. Routh's procedure, which is Lagrangian in nature,  deals with the so-called cyclic or ignorable variables (variables on which the Lagrangian does not depend explicitily): cyclic variables lead to conserved momenta, and these allow to construct a reduced Lagrangian function, nowadays known as the Routhian. The Euler-Lagrange equations for the Routhian involve fewer variables than the un-reduced Euler-Lagrange equations, and the solutions of the dynamical equations for the Routhian, together with a precribed value of momenta for the un-reduced system, can be  used to reconstruct solutions of the original Lagrangian system. 

Since then, a number of papers have been devoted to the geometrization  and generalization of this reduction technique (e.g. \cite{TomMike,GU,quasi}) but, to the best of our knowledge, the hybrid analogue has not been fully discussed in the literature. An hybrid scheme for Routh reduction for autonomous hybrid Lagrangian systems with cyclic variables is found in~\cite{AmesL}; this technique (or a variant of it, the so-called functional Routh reduction) has been successfully applied to the study control strategies of certain bipedal walkers, see e.g. \cite{ames2007,Gregg08}. This work attempt to go one step further and discuss Routh reduction for non-autonomous hybrid systems. We remark that some of the techniques outlined here have been recently applied to the study of periodic orbits in reduced hybrid Lagrangian systems~\cite{BlClCo, LeoEmma,CoClBl,EmmaPhd}.

The material is organized as follows. Section $2$ contains some preliminary results on the geometry of time-dependent Lagrangian systems. Section $3$ discusses the notion of hybrid time-dependent mechanical system with symmetry. The reduction scheme is proposed in Section $4$. Finally, Section $5$ contains an illustrative example: the time-dependent system described by a rough billiard with moving walls.

\section{Time-dependent hybrid systems and symmetries} We start recalling some basic facts about time-dependant mechanics. Our starting point is a time-dependent Lagrangian $L:\mathbb{R}\times TQ\to\mathbb{R}$; we will denote by $\mathbb{F} L\colon \mathbb{R}\times TQ\to \mathbb{R}\times T^*Q$ the Legendre transformation associated with $L$, which is the map $(t,q,\dot q)\mapsto (t,q,p=\partial L/\partial \dot q)$.

We will assume that the Lagrangian is hyperregular, i.e. that $\mathbb{F} L$ is a diffeomorphism between $\mathbb{R}\times TQ$ and $\mathbb{R}\times T^*Q$ (this is always the case for mechanical Lagrangians). One can then work out the velocities $\dot q$ in terms of $(t,q,p)$ using the inverse of $\mathbb{F} L$ and define the Hamiltonian function  $H\colon \mathbb{R}\times T^*Q\to \mathbb{R}$ as
\[
H(t,q,p)=\langle p, \dot q(t,q,p)\rangle - L(q,\dot q(t,q,p)).
\]

Among the several geometric approaches to non-autonomous mechanics~\cite{Setting}, we will be using one based on the notion of cosymplectic manifold (see Appendix for details). We first describe the procedure for Hamiltonian mechanics. To describe the dynamics of a non-autonomous Hamiltonian system, one begins by considering the manifold $\mathbb{R}\times T^*Q$ equipped with the canonical cosymplectic structure 
\[
\Omega=dq\wedge dp,\quad \eta=dt, 
\]
with Reeb vector field $\mathbf{T}=\partial_t$. Given a Hamiltonian $H(t,q,p)$, the \textit{Hamiltonian vector field} $X_H$ is the vector field on $\mathbb{R}\times T^*Q$ defined by 
\[
i_{X_H}\Omega=dH-\mathbf{T}(H)\eta,\quad i_{X_H}\eta=0. 
\]
The evolution vector field corresponding to the Hamiltonian $H$, denoted by $Z_H$, is defined by 
\[
Z_H=\mathbf{T}+X_H=\frac{\partial}{\partial t}+\frac{\partial H}{\partial p}\frac{\partial }{\partial q}-\frac{\partial H}{\partial q}\frac{\partial }{\partial p},
\]
and its integral curves are solutions of the Hamilton equations $\dot q=\partial_pH$, $\dot p=-\partial_qH$, where the ``$\partial_t$'' component of $Z_H$ sets the evolution parameter of the solutions to be the ``time'' $t$. 

In the Lagrangian picture, under the assumption of hyperregularity discussed above, a similar procedure can be used. The manifold $\mathbb{R}\times TQ$ can be endowed with the following cosymplectic structure
\begin{equation}\label{eq:2}
\Omega_L=\mathbb{F} L^*(dq\wedge dp)=dq\wedge d\left(\frac{\partial L}{\partial\dot{q}}\right),\quad \eta=dt,
\end{equation}
which depends on the Lagrangian (note that we will be using the same symbol $\eta=dt$ for two different 1-forms on different manifolds). One then constructs the \emph{energy} function $E_L\colon \mathbb{R}\times TQ\to \mathbb{R}$ given by 
\[
E_L(t,q,\dot q)=\langle \mathbb{F} L(t,q,\dot q),\dot q\rangle - L(t,q,\dot q),  
\]
and obtains the Hamiltonian vector field associated to the cosymplectic structure~\eqref{eq:2} and Hamiltonian $E_L$. This leads to an evolution vector field, that we denote $Z_L$, given by
\[
Z_L=\frac{\partial}{\partial t}+\dot q \frac{\partial }{\partial q}+ \Gamma(t,q,\dot q)\frac{\partial }{\partial \dot{q}},
\]
where $\Gamma(t,q,\dot q)$ is obtained by isolating $\ddot q$  from the standard Euler-Lagrange equations
\[
\frac{d}{dt}\left(\frac{\partial L}{\partial\dot{q}}\right)-\frac{\partial L}{\partial{q}}=0. 
\]

Finally, the well-known equivalence between the Lagrangian and Hamiltonian dynamics in the hyperregular case is achieved via $\mathbb{F} L$.

\begin{proposition}\label{Proposition1} The tangent map of $\mathbb{F} L$ maps $Z_L$ onto $Z_H$, i.e.  $(T\, \mathbb{F} L) (Z_L)=X_H$. In particular, the flow of $Z_L$ is mapped onto the flow of $Z_H$. 
\end{proposition}

\textbf{Proof:} The evolution vector field $Z_H$ is characterized by $i_{X_H}\Omega=dH-\mathbf{T}(H)\eta$ and $i_{X_H}\eta=1$. Now:
\begin{align*}
(\mathbb{F} L)^{*}(i_{Z_{H}}\Omega)&=(\mathbb{F} L)^{*}(dH-\mathbf{T}(H)\eta)=(\mathbb{F} L)^{*}(dH)-(\mathbb{F} L)^{*}(\mathbf{T}(H)\eta)\\
&=d((\mathbb{F} L)^{*}H)-\mathbf{T}(\mathbb{F} L(H))\eta=d(E_{L})-\mathbf{T}(E_{L})\eta\\
&=i_{Z_{L}}\Omega_{L}.
\end{align*}
This means $i_{Z_{L}}\Omega_{L}=(\mathbb{F} L)^{*}(i_{Z_{H}}\Omega)=i_{(\mathbb{F} L^{-1})_{*}Z_{H}}(\mathbb{F} L^{*}\Omega)=i_{(\mathbb{F} L^{-1})_{*}Z_{H}}\Omega_{L}$. In a similar way one shows that $i_{Z_L}\eta=i_{(\mathbb{F} L^{-1})_{*}Z_{H}}\eta$. This implies $Z_{L}=(\mathbb{F} L^{-1})_{*}Z_{H}$, that is, $(\mathbb{F} L)_{*}Z_{L}=Z_{H}.$ \hfill$\square$

\section{Simple Hybrid time-dependent Mechanical Systems and Symmetries}

Roughly speaking, the term \emph{hybrid system} refers to a dynamical system which exhibits both continuous and discrete behavior. In the literature, one finds slightly different definitions of hybrid system depending on the specific class of applications of interest. For simplicity, and following~\cite{AmesH} and \cite{AmesL}, we will restrict ourselves to the so-called simple hybrid mechanical systems in Lagrangian form. We will extend the definition in order to include time dependence on both the Lagrangian and the switching surface (see below).

\begin{definition} A \emph{simple hybrid time-dependent Lagrangian system} is a tuple $\mathscr{L}=(Q,L,S,R)$  where
	\begin{itemize}
		\item[(i)] $Q$ is a differentiable manifold,
		\item[(ii)] $L\colon \mathbb{R}\times TQ\to \mathbb{R}$ is a time-dependent Lagrangian (recall that we will always assume hyperregularity),
		\item[(iii)] $S$ is an embedded submanifold of $\mathbb{R}\times TQ$ with co-dimension one called the \emph{switching surface} (sometimes referred to as the \emph{guard}),
		\item[(iv)] $R\colon S\to \mathbb{R}\times TQ$ is a smooth map called the \emph{reset map} (often called the impact map).
	\end{itemize}
	The triple $(Q,S,R)$ alone is referred to as an \emph{hybrid manifold}.
\end{definition}

In a similar way one defines a \emph{hybrid time-dependent Hamiltonian system} as a tuple
$\mathscr{H}=(Q,H,S_{H},R_{H})$, where $H\colon\mathbb{R}\times T^*Q\to\mathbb{R}$ is a time-dependent Hamiltonian function and the elements of the hybrid manifold are now defined on $T^{*}Q$ instead of $TQ$.

We have recalled in Proposition \ref{Proposition1} that in the hyperregular case there is an equivalence between the Lagrangian and the Hamiltonian descriptions of a mechanical system.  We will now extend this equivalence to the hybrid setting (among other things, this clarifies  the relation between the results in \cite{AmesH} and \cite{AmesL}). We need the following definition:

\begin{definition}\label{def:flow} A \emph{hybrid flow} for $\mathscr{L}$ is a tuple $\chi^{\mathscr{L}}=(\Lambda,\mathcal{J},\mathscr{C})$, where
	\begin{itemize}
		\item $\Lambda=\{0,1,2,...\}\subseteq \mathbb{N}$ is a finite (or infinite) indexing set,
		\item $\mathcal{J}=\{I_{i}\}_{i\in \Lambda}$ a set of intervales, called hybrid intervals where $I_{i}=[\tau_{i},\tau_{i+1}]$ if $i, i+1\in \Lambda$ and $I_{N-1}=[\tau_{N-1},\tau_{N}]$ or $[\tau_{N-1},\tau_{N})$ or $[\tau_{N-1},\infty)$ if $|\Lambda|=N$, $N$ finite, with $\tau_{i},\tau_{i+1},\tau_{N}\in \mathbb{R}$ and $\tau_{i}\leq \tau_{i+1}$,
		\item $\mathscr{C}=\{c_{i}\}_{i\in \Lambda}$ is a collection of solutions for the vector field $Z_L$ specifying the continous-time dynamics, i.e., $\dot{c_{i}}=Z_L(c_{i}(t))$ for all $i\in \Lambda$,
		and such that for each $i,i+1\in \Lambda$,
		\begin{enumerate}
			\item[(i)] \quad $c_{i}(\tau_{i+1})\in S$,
			\item[(ii)]\quad $R(c_{i}(\tau_{i+1}))=c_{i+1}(\tau_{i+1})$.
		\end{enumerate} 
	\end{itemize}
\end{definition}

Analogously, one defines the notion of hybrid flow $\chi^{\mathscr{H}}$ for a hybrid time-dependent Hamiltonian system $\mathscr{H}$. The relation between both the Lagrangian and the Hamiltonian hybrid flows is given  by the following result, where for clarity in the exposition we will use the notation $\mathscr{L}=(Q,L,S,R)$, $\mathscr{H}=(Q,H,S_H,R_H)$. 
\begin{proposition}\label{Proposition2}
	If $\chi^{\mathscr{L}}=(\Lambda,\mathcal{J},\mathscr{C})$ is a hybrid flow for $\mathscr{L}$, $S_H=\mathbb{F} L(S)$, and $R_H$ is defined in such a way that $\mathbb{F} L\circ R=R_H\circ \mathbb{F} L\mid_{S}$, then $\chi^{\mathscr{H}}=(\Lambda,\mathcal{J},(\mathbb{F} L)(\mathscr{C}))$ with $(\mathbb{F} L)(\mathscr{C})=\{(\mathbb{F} L)(c_{i})\}_{i\in \Lambda}$.
\end{proposition}

\textbf{Proof:} By Proposition \ref{Proposition1}, if $c_i(t)$ is an integral curve of $Z_{L}$, $\tilde c_i(t)=(\mathbb{F} L\circ c_i)(t)$ is an integral curve for $X_{H}$. In this way, if we consider a solution $c_0(t)$  with initial value $c_0=(q_0,\dot q_0)$ defined on $[\tau_0,\tau_1]$, then $\tilde c_0(t)$ is a solution with initial value $\tilde c_0=(q_0,p_0)$ defined on $[\tau_0,\tau_1]$. Likewise for a solution $c_1(t)$ defined on $[\tau_1,\tau_2]$, we get a corresponding solution $\tilde c_1(t)$ defined on the same hybrid interval $[\tau_1,\tau_2]$. Proceeding inductively, one finds $c_i(t)$ defined on  $[\tau_i,\tau_{i+1}]$. 	It only remains to check that $\tilde c_i(t)$ satisfies $\tilde c_i(\tau_{i+1})\in S_{H}$ and $R_{H}(\tilde c_{i}(\tau_{i+1}))=\tilde c_{i+1}(\tau_{i+1})$, but using the properties of $\mathbb{F} L$, we have that,
\begin{itemize}
	\item[(i)] $\tilde c_i(\tau_{i+1})=(\mathbb{F} L\circ c_{i})(\tau_{i+1})=\mathbb{F} L(c_i(\tau_{i+1}))$ and given that $c_i(\tau_{i+1})\in S$ then $\tilde c_i(\tau_{i+1})\in S_{H}.$
	\item[(ii)]  $R_H(\tilde c_{i}(\tau_{i+1}))=R_H\circ \mathbb{F} L\circ c_{i}(\tau_{i+1})=\mathbb{F} L\circ R\circ c_{i}(\tau_{i+1})=\mathbb{F} L\circ c_{i+1}(\tau_{i+1})=\tilde c_{i+1}(\tau_{i+1})$.\hfill$\square$
\end{itemize}

Let $\mathscr{L}=(Q,L,S,R)$ be a simple hybrid time-dependent Lagrangian system. The starting point for symmetry reduction is a Lie group action $\psi\colon G\times Q\to Q$ of some Lie group $G$ on the manifold $Q$. We will assume that all the actions satisfy some regularity conditions as to do reduction (for instance, one can consider free and proper actions).

There is a natural lift $\Psi$ of the action $\psi$ to the space $\mathbb{R}\times T^*Q$, the cotangent lift action, defined by $\Psi_g=T^*\psi_{g^{-1}}$. It enjoys the following properties~\cite{Albert}:
\begin{itemize}
	\item $\Psi$ is a cosymplectic action, meaning that $\Psi_g^*\Omega=\Omega$ and $\Psi^*\eta=\eta$.
	\item It admits an $\hbox{Ad}^*$-equivariant momentum map 
	$
	J\colon \mathbb{R}\times T^*Q\to \mathfrak{g}^*$ given by
	\[
	\langle J(t,q,p), \xi\rangle=\langle p, \xi_Q\rangle,\quad \forall \xi\in\mathfrak{g},
	\]
	where $\xi_Q(q)=d(\psi_{\exp(t\xi)}q)/dt$ is the infinitesimal generator of  $\xi\in \mathfrak{g}$.
\end{itemize}
Likewise, $\Psi^{TQ}$ denotes the tangent lift action on $\mathbb{R}\times TQ$, defined by $\Psi^{TQ}_g=\psi_{g}(q,\dot q)$.

To perform a hybrid reduction one needs to impose some compatibility conditions between the action and the hybrid system (see e.g.~\cite{AmesL}). By an \emph{hybrid action} on the simple hybrid time-dependent Lagrangian system $\mathscr{L}=(Q,L,S,R)$  we mean a Lie group  action $\psi\colon G\times Q\to Q$ such that
\begin{itemize}
	\item $L$ is invariant under $\Psi^{TQ}$, i.e. $L\circ \Psi^{TQ}=L$.
	\item $\Psi^{TQ}$ restricts to an action of $G$ on $S$.
	\item $R$ is equivariant with respect to the previous action, namely $R\circ \Psi^{TQ}_g\mid_S=\Psi^{TQ}_g\circ R$.
\end{itemize}
Recall that $\Psi^{TQ}$ admits an $\hbox{Ad}^*$-equivariant momentum map $J_L:\mathbb{R}\times TQ\to\mathfrak{g}^{*}$ given by $J_L=J\circ\mathbb{F} L$. This follows directly from the invariance of $L$, since it implies that $\mathbb{F} L$ is an equivariant diffeomorphism, i.e. $\mathbb{F} L\circ \Psi^{TQ}_g=\Psi_g\circ \mathbb{F} L$
\[
\mathbb{F} L\circ \Psi^{TQ}_g=\Psi_g\circ \mathbb{F} L. 
\]

The hybrid equivalent of momentum map is the notion of hybrid momentum map introduced in  \cite{AmesH}. In the case of $\mathbb{R}\times T^*Q$, $J$ is an  \emph{hybrid momentum map} if the  diagram
\begin{equation}\label{diag1}
\begin{tikzcd}[column sep=1.5cm, row
sep=1.2cm]
& \mathfrak{g}^* &\\
\mathbb{R}\times T^*Q \arrow[ur,"J"] & S \arrow[u,"J\mid_S"] \arrow[l,swap,hook',"i"] \arrow[r,"R"] & \mathbb{R}\times T^*Q  \arrow[ul, swap,"J"]
\end{tikzcd} 
\end{equation}
commutes, where $i\colon S\hookrightarrow \mathbb{R}\times T^{*}Q$ denotes the canonical inclusion. 

\section{Reduction by symmetries of simple hybrid time-dependent Lagrangian system} 

Consider a simple hybrid time-dependent Lagrangian system $\mathscr{L}=(Q,L,S,R)$  equipped with an hybrid action $\psi$. We begin analyzing the reduction of the associated hybrid time-dependent Hamiltonian system $\mathscr{H}=(Q,H,S_H,R_H)$ discussed in Proposition \ref{Proposition2}. 

Consider a \textit{hybrid regular value} $\mu\in\mathfrak{g}^*$ of $J\colon \mathbb{R}\times T^*Q\to\mathfrak{g}^*$, which means that $\mu$ is a regular value of both $J$ and $J\mid_S\colon S\to\mathfrak{g}^*$. When we combine this definition with the 
commuting diagram~\eqref{diag1}, we obtain that the following diagram
\[
\begin{tikzcd}[column sep=1.3cm, row
sep=1.2cm]
J^{-1}(\mu)\arrow[d,hook'] &  J\mid_S^{-1}(\mu)\arrow[r,"R_H\mid_S"] \arrow[l,swap,hook',"i"]\arrow[d,hook'] & J^{-1}(\mu)\arrow[d,hook']\\
\mathbb{R}\times T^*Q  & S \arrow[l,swap,hook',"i"] \arrow[r,"R_H"] & \mathbb{R}\times T^*Q  
\end{tikzcd} 
\] commutes, where $J^{-1}(\mu)$ and $J\mid_{S}^{-1}(\mu)$ are embedded submanifolds of $\mathbb{R}\times T^{*}Q$ and $S$, respectively.

We can apply a hybrid analog of the cosymplectic reduction Theorem \cite{Albert} to the hybrid time-dependent Hamiltonian system $\mathscr{H}$ (see Appendix). Note that, since $L$ is invariant under $\Psi^{TQ}$, so is the Hamiltonian $H$ under $\Psi^{T^{*}Q}$. The main conclusions are:
\begin{enumerate}
	\item[(i)] The reduced space $J^{-1}(\mu)/G_\mu$ (with $G_\mu$ the isotropy group of $\mu$ under the coadjoint action) is a cosymplectic manifold, and the reduced cosymplectic structure $(\eta_\mu,\Omega_\mu)$ is characterized in terms of the submersion $\pi_\mu\colon J^{-1}(\mu)\to J^{-1}(\mu)/G_\mu$ and the inclussion $i_\mu\colon J^{-1}(\mu)\hookrightarrow \mathbb{R}\times T^*Q$ by means of the relations $\pi_\mu^*\eta_\mu=i_\mu^*\eta$ and $\pi_\mu^*\Omega_\mu=i_\mu^*\Omega$. 
	\item[(ii)] If we denote by $H_\mu$ the reduction of $H\mid_{J^{-1}(\mu)}$ to $J^{-1}(\mu)/G_\mu$, the evolution vector field $Z_H$ projects onto $Z_{H_\mu}$. This is the second part of the Cosymplectic reduction Theorem of Albert ~\cite{Albert} (see Appendix).
	\item[(iii)] $J\mid_S^{-1}(\mu)\subset S_H$ is $G_\mu$-invariant and hence  reduces to a submanifold of the reduced space which we denote $(S_H)_{\mu}\subset J^{-1}(\mu)/G_\mu$.
	\item[(iv)] Again, using invariance $R_H$ reduces to a map $(R_H)_{\mu}\colon (S_H)_{\mu}\to J^{-1}(\mu)/G_\mu$.
\end{enumerate}

The reduction picture in the Lagrangian side can now be obtained from the Hamiltonian one by adapting the scheme developed in~\cite{quasi} to the cosymplectic setting. The key idea is that, since $L$ is invarian, the Legendre transformation $\mathbb{F} L$ is a diffeomorphism such that:
\begin{itemize}
	\item It is equivariant with respect to $\Psi^{TQ}$ and $\Psi^{T^{*}Q}$, 
	\item Preserves the level sets of the momentum map, that is, $\mathbb{F} L (J_L^{-1}(\mu))=J^{-1}(\mu),$
	\item Relates both cosymplectic structures, that is, $(\mathbb{F} L)^*\Omega=\Omega_L$ and $(\mathbb{F} L)^*\eta=\eta$ ($\mathbb{F} L$ is sometimes referred to as a \emph{cosymplectomorphism}).
\end{itemize}
It follows that the map $\mathbb{F} L$ reduces to a cosymplectomorphism $(\mathbb{F} L)_{\text{red}}$ between the reduced spaces. Therefore we get the following commutative diagram of hybrid manifolds:
\[
\begin{tikzcd}[column sep=1.6cm, row
sep=1.2cm]
(\mathbb{R}\times TQ,S,R)  \arrow[d,swap,"\text{Red.}"] \arrow[r,"\mathbb{F} L"] & (\mathbb{R}\times T^*Q,S_H,R_H) \arrow[d,"\text{Red.}"]\\
(J_L^{-1}(\mu)/G_\mu,S_\mu,R_\mu) \arrow[r,"(\mathbb{F} L)_{\text{red}}"] & (J^{-1}(\mu)/G_\mu,(S_H)_{\mu},(R_H)_{\mu}) 
\end{tikzcd} 
\]

We now make use of a principal connection on the bundle $Q\to Q/G$ to make some further identifications. Let $\mathcal{A}\colon TQ\to\mathfrak{g}^*$ be the connection one form, and let us denote by $\mathcal{A}_\mu(\cdot)=\langle \mu,\mathcal{A}(\cdot)\rangle$ the 1-form on $Q$ obtained by contraction with $\mu\in\mathfrak{g}^*$. Building on the well-known results on cotangent bundle reduction it is possible to show that there is an identification
\begin{equation}\label{eq:ide}
J^{-1}(\mu)/G_\mu\simeq \mathbb{R}\times \left(T^*(Q/G)\times_{Q/G}Q/G_\mu\right). 
\end{equation}
This identification is a symplectomorphism when we endow the space on the right hand side of~\eqref{eq:ide}  with the symplectic structures ${\rm pr}_1^*\Omega_{Q/G}+{\rm pr}_2^*\mathcal{B}_\mu$, where ${\rm pr}_1$ and ${\rm pr}_2$ are the canonical projections, and $\mathcal{B}_\mu$ is the so-called magnetic term, obtained from the reduction of $d\mathcal{A}_\mu$ to $Q/G_\mu$. For details, see~\cite{quasi,Hstages}.

For the Lagrangian side, one needs a further regularity condition, sometimes referred to as \emph{$G$-regularity}, which is satisfied by mechanical Lagrangians. Precisely, one has the following definition~\cite{Routhstages} (for an alternative, equivalent definition, see~\cite{quasi}):

\begin{definition} Let $L$ be an invariant Lagrangian on $TQ$ and denote by $\xi_Q$ the infinitesimal generator for the associated action. We say that $L$ is \emph{$G$-regular} if, for each $v_q\in TQ$, the map
\begin{align*}
\mathcal{J}_L^{v_q}:\mathfrak{g}&\to \mathfrak{g}^*,\\
\xi&\mapsto J_L\left(v_q + \xi_Q(q)\right),
\end{align*}
is a diffeomorphism.
\end{definition}
In a nutshell, $G$-regularity amounts to regularity ``with respect to the group variables''. From now on we will assume that the Lagrangian is $G$-regular. In this case, there is an identification 
\begin{equation*}
J_L^{-1}(\mu)/G_\mu\simeq \mathbb{R}\times \left(T(Q/G)\times_{Q/G}Q/G_\mu\right). 
\end{equation*}
It is possible to interpret the reduced dynamics on this space as being the Lagrangian dynamics of some regular Lagrangian subjected to a gyroscopic force (arising from the magnetic term) if one works in the more general class of \emph{magnetic Lagrangians}~\cite{GT}, which in the present situation should be extended to include time-dependent Lagrangians. The so-called magnetic Lagrangian systems are a wide class of Lagrangian systems on which the Lagrangian function might not depend on some of the velocities, and which may as well include a force term given by a 2-form. The framework of magnetic Lagrangian systems is very convenient when carrying out Routh reduction, since the reduced system is not, in general, a standard Lagrangian system. We remark that Routh reduction has been extended to magnetic Lagrangian system, and this permits to carry out Routh reduction by stages~\cite{Routhstages}. 

The function which plays the role of the reduced Lagrangian is the \emph{Routhian}, and it is defined as (the reduction of) the $G_\mu$-invariant function
\begin{equation}\label{eq:Routhiandefinition}
L_\mu=L-\mathcal{A}_\mu
\end{equation}
restricted to $J_L^{-1}(\mu)$. The next diagram summarizes the situation:
\[
\begin{tikzcd}[column sep=1.6cm, row
sep=1.2cm]
\mathbb{R}\times TQ  \arrow[d,swap,"\text{Red.}"] \arrow[r,"\mathbb{F} L"] & \mathbb{R}\times T^*Q \arrow[d,"\text{Red.}"]\\
 \mathbb{R}\times \left(T(Q/G)\times_{Q/G}Q/G_\mu\right) \arrow[r,"\mathbb{F} L_\mu"] &  \mathbb{R}\times \left(T^*(Q/G)\times_{Q/G}Q/G_\mu\right)
\end{tikzcd} 
\]
We will now focus on the particular case of cyclic coordinates, which is the one in encountered in the reeferences motivating this note. More details and examples on the general situation can be found in~\cite{EmmaPhd}.

\vspace{.5em}\noindent\textbf{Cyclic coordinates in simple hybrid time-dependent Lagrangian system} The case $G=\mathbb{S}^1$  corresponds to the notion of cyclic coordinates (the case $G=\mathbb{R}$ is analogous; if $G$ is a product one can iterate the procedure). The reduced space  $J_L^{-1}(\mu)/G_\mu$ can be identified with $\mathbb{R}\times T(Q/\mathbb{S}^1)$ and the reduced dynamics is Lagrangian with respect to the reduced Lagrangian $L_\mu=L-\mathcal{A}_\mu$ on $\mathbb{R}\times T(Q/\mathbb{S}^1)$. The reduced switching $S_\mu$ can be identified with a submanifold of $\mathbb{R}\times T(Q/\mathbb{S}^1)$ and the reset map is identified with a map $R_\mu\colon S_\mu\to \mathbb{R}\times T(Q/\mathbb{S}^1)$. We will use the same notations for both of them.

A case of special interest with regards to applications is when $Q=\mathbb{S}^1\times M$, where $M$ is called the \emph{shape space} and the action is simply $(\theta,x)\mapsto (\theta+\alpha,x)$. This is often the situation when dealing with simple models of bipedal walkers, see e.g.~\cite{ames2007,Gregg08}.  From now on, we will assume we work in this setting. While this is indeed a strong assumption, it is always the case locally, so as long as it applies to the domain of interest of an specific problem the procedure applies. The Lagrangian has a cyclic coordinate $\theta$, i.e. $L$ is a function of the form $L(t,\dot \theta,x,\dot x)$. The conservation of the momentum map $J=\mu$ reads $\partial L/\partial \dot\theta=\mu$, and one can use this relation to express $\dot\theta$ as a function of the remaining -non cyclic- coordinates and their velocities, and the prescribed regular value of the momentum map $\mu$. We point out that it is at this stage that $G$-regularity of $L$ is used: it guarantees that $\dot\theta$ can be worked out in terms of $x$, $\dot x$ and $\mu$. If one chooses the cannonical flat connection on $Q\to Q/\mathbb{S}^1=M$, then the Routhian can be computed as
\begin{equation}\label{eq:Rclas}
L_\mu(t,x,\dot x)=\left[L(t,\dot \theta,x,\dot x)-\mu\dot\theta\right]\Big{|}_{\dot\theta=\dot\theta(t,x,\dot x,\mu)},
\end{equation}
where the notation means that we have everywhere expressed $\dot\theta$ as a function of $(t,x,\dot x,\mu)$. Note that~\eqref{eq:Rclas} is the classical definition of the Routhian~\cite{Pars}. Besides, since the connection is flat, one has no force term in the reduced dynamics. 

Let us first consider the case in which the momentum map is preserved in the collisions with the switching surface (elastic case). We then have:

\begin{proposition}\label{prop:main} In the situation above:
	\begin{enumerate}
		\item[(i)] Any solution of $\mathscr{L}=(Q,L,S,R)$ with momentum $\mu$ projects onto a solution of $\mathscr{L}_\mu=(Q/\mathbb{S}^1,L_\mu,S_\mu,R_\mu)$.
		\item[(ii)] Any solution of $\mathscr{L}_\mu=(Q/\mathbb{S}^1,L_\mu,S_\mu,R_\mu)$ is the projection of a solution of $\mathscr{L}=(Q,L,S,R)$ with momentum $\mu$.
	\end{enumerate}
\end{proposition}

Collisions with the switching surface will, in general, modify the value of the momentum map (non-elastic case).  Therefore, if $\mathcal{J}=\{I_{i}\}_{i\in \Lambda}$ is the hybrid interval (see Definition~\ref{def:flow}), the Routhian has to be defined in each $I_i$ taking into account the value of the momentum $\mu_i$ after the collision at time $\tau_i$. Note that this also has an influence in the way the reset map $R$ and the switching $S$ are reduced. Let us denote: (1) $\mu_i$ the momentum of the system in $I_i=[\tau_i,\tau_{i+1}]$, (2) $R_{\mu_i}$ the reduction of $R\mid_{J^{-1}(\mu_i)}$, and (3) $S_{\mu_i}$ the reduction of $J\mid_S^{-1}(\mu_i)$, There is a sequence of reduced simple hybrid time-dependent Lagrangian systems (``coll'' stands for collision):
\[
\begin{tikzcd}[column sep=.5cm, row
sep=.7cm]
{[\tau_0,\tau_1]} \arrow[d,swap,"\text{coll.}"]\arrow[r,"\text{Red.}"] & (Q/\mathbb{S}^1,L_{\mu_0},S_{\mu_0},R_{\mu_0}) \arrow[d,"\text{coll.}"]\\
{[\tau_1,\tau_2]} \arrow[d,swap,"\text{coll.}"]\arrow[r,"\text{Red.}"] & (Q/\mathbb{S}^1,L_{\mu_1},S_{\mu_1},R_{\mu_1}) \arrow[d,"\text{coll.}"]\\
(\dots) \arrow[r,"\text{Red.}"] & (\dots)
\end{tikzcd} 
\]

The fact that the momentum will, in general, change with the collisions makes the reconstruction procedure more challenging. If one wishes, as usual, to use a reduced solution to reconstruct the original dynamics, one needs to compute the reduced hybrid data after each collision. This means that once the reduced solution has been obtained between two collison events, say at $t=\tau_n$ and $t=\tau_{n+1}$, one should reconstruct this solution to obtain the new momentum after the collision at $\tau_{n+1}$ and use this new momentum to build a new reduced hybrid system whose solution should be obtained until the next collision eventy at $\tau_{n+2}$, and so on. As usual, the reconstruction procedure from the reduced hybrid flow to the hybrid flow involves an integration at each stage in the previous diagram of the cyclic variable using the solution of the reduced simple hybrid time-dependent Lagrangian system. Essentially, this accounts to imposing the momentum constraint on the reconstructed solution.

\begin{remark} Proposition~\ref{prop:main} is easily adapted to more general scenarios. If $Q$ is not a product, one needs to compute the Routhian using the general expression~\eqref{eq:Routhiandefinition} and consider the reduced simple hybrid time-dependent Lagrangian system with a force term. If $G$ is non-Abelian one can use the class of magnetic Lagrangian systems to describe the reduced dynamics. Coordinate formulae for the reduced dynamics and reconstruction in this more general case can be obtained using the techniques in~\cite{TomMike}.
\end{remark}

\section{Example: a rough billiard with moving walls}

\begin{figure}[h]
	\centering\includegraphics{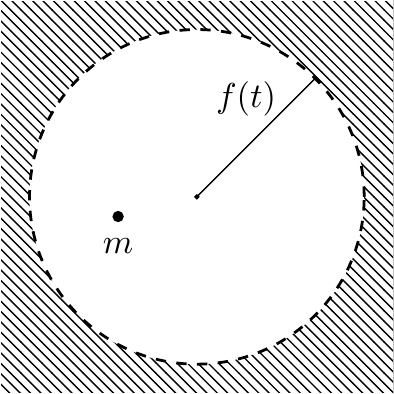}
	\caption{A ``billiard'' with moving walls}\label{fig:1}
\end{figure}

Consider a particle of mass $m$ in the plane which is free to move inside the surface defined by a circle whose radius varies in time (Figure~\ref{fig:1}) according to a given function $f(t)$, i.e. $x^2+y^2=f(t).$ 
The surface of the ``billiard'' is assumed to be rough in such a way that the friction is proportional to the velocity (the dissipation is of \emph{Rayleigh} type, see e.g. \cite{Bloch}). This example falls in the category of mechanical systems with one-sided constraints, see e.g.~\cite{cortes, Constraints, LC} for alternative approaches.

The equations of motion for the particle off the boundary are 
\begin{equation}\label{eq:ex1}
m\ddot x=-c \dot x,\quad  m\ddot y=-c \dot y.
\end{equation}
for some constant $c>0$ (a dissipation coefficient). To fit these equations into Lagrangian form, one considers the time-dependent Lagrangian $L:\mathbb{R}\times T\mathbb{R}^{2}\to\mathbb{R}$ given by 
\[
L(t,x,y,\dot{x},\dot{y})=\exp{\left(\frac{c}{m}t\right)}\left[\frac{1}{2}m(\dot x^2+\dot y^2)\right].
\]
It is easy to check that the Euler-Lagrange equations for $L$ give the desired equations of motion. The guard is the subset of $\mathbb{R}\times T\mathbb{R}^2\simeq\mathbb{R}\times\mathbb{R}^{2}\times\mathbb{R}^{2}$ given by
\[
S=(\mathbb{R}\times TQ)\cap \{x^2+y^2=f(t), (\dot x,\dot y)\cdot (x,y)\geq 0 \}. 
\]
This set $S$ describes the situation in which the particle hits the moving boundary while heading ``outwards'' the billiard. For simplicity in the definition of the switching surface, we assume that $f(t)$ is increasing: this guarantees the particle only hits the boundary when the boundary is also moving outwards. Under the assumption of an elastic collision, the reset map
\[
(t,x,y,\dot x^{-},\dot y^{-})\mapsto (t,x,y,\dot x^{+},\dot y^{+})
\]
is given by~\cite{Constraints}:
\begin{equation*}
\dot x^{+}=\dot x^{-} + \frac{\dot f(t)-2 (x\dot x^{-}+y\dot y^{-})}{f(t)} x,\qquad
\dot y^{+}=\dot y^{-} + \frac{\dot f(t)-2 (x\dot x^{-}+y\dot y^{-})}{f(t)} y.
\end{equation*}

Note that, if $f(t)={\rm constant}$, the reset map agrees with the one given for a constraint function $h\colon Q\to \mathbb{R}$ in~\cite{AmesL} in the case of a perfectly elastic impact. Introducing polar coordinates the Lagrangian becomes
\[
L(t,r,\theta,\dot{r},\dot{\theta})=\exp{\left(\frac{c}{m}t\right)}\frac{m}{2}(\dot r^2+ r^2\dot \theta^2)
\]
for which $\theta$ is a cyclic coordinate. In this case  $J_L$ is time dependent and represents a damped angular momentum ($m r^2 \dot\theta$ decays exponentially in time as a result of the friction): 
\begin{equation}\label{eq:Mom2}
J_L(t,r,\dot r,\theta,\dot \theta)= \exp{\left(\frac{c}{m}t\right)} mr^2\dot\theta. 
\end{equation}

A computation reveals that the reset map, in polar coordinates, takes the form (observe that $2(x\dot x^{-}+y\dot y^{-})$ is nothing but  $2 r \dot r^{-}$):
\begin{align}\label{eq:resetpolar}
(\dot{r}^{+})^{2}&=(\dot{r}^{-})^2+\nonumber\\
&+\frac{r}{f(t)}(\dot{f}(t)-2r\dot{r}^{-})\left(2\dot{r}^{-}+\frac{(\dot{f}(t)-2r\dot{r}^{-})r}{f(t)}\right),\\
\dot{\theta}^{+}&=\dot{\theta}^{-}.\nonumber 
\end{align}
It is understood that the ``minus'' square root is taken in $\dot{r}^{+}$ (the particle bounces on the boundary after the collision). The assumption of elastic collision implies, in particular, that the momentum map is preserved. This is clear since $r$ and $\dot\theta$ do not change with the collision~\eqref{eq:resetpolar}.

\begin{figure}[h]
	\centering\includegraphics[height=5.7cm]{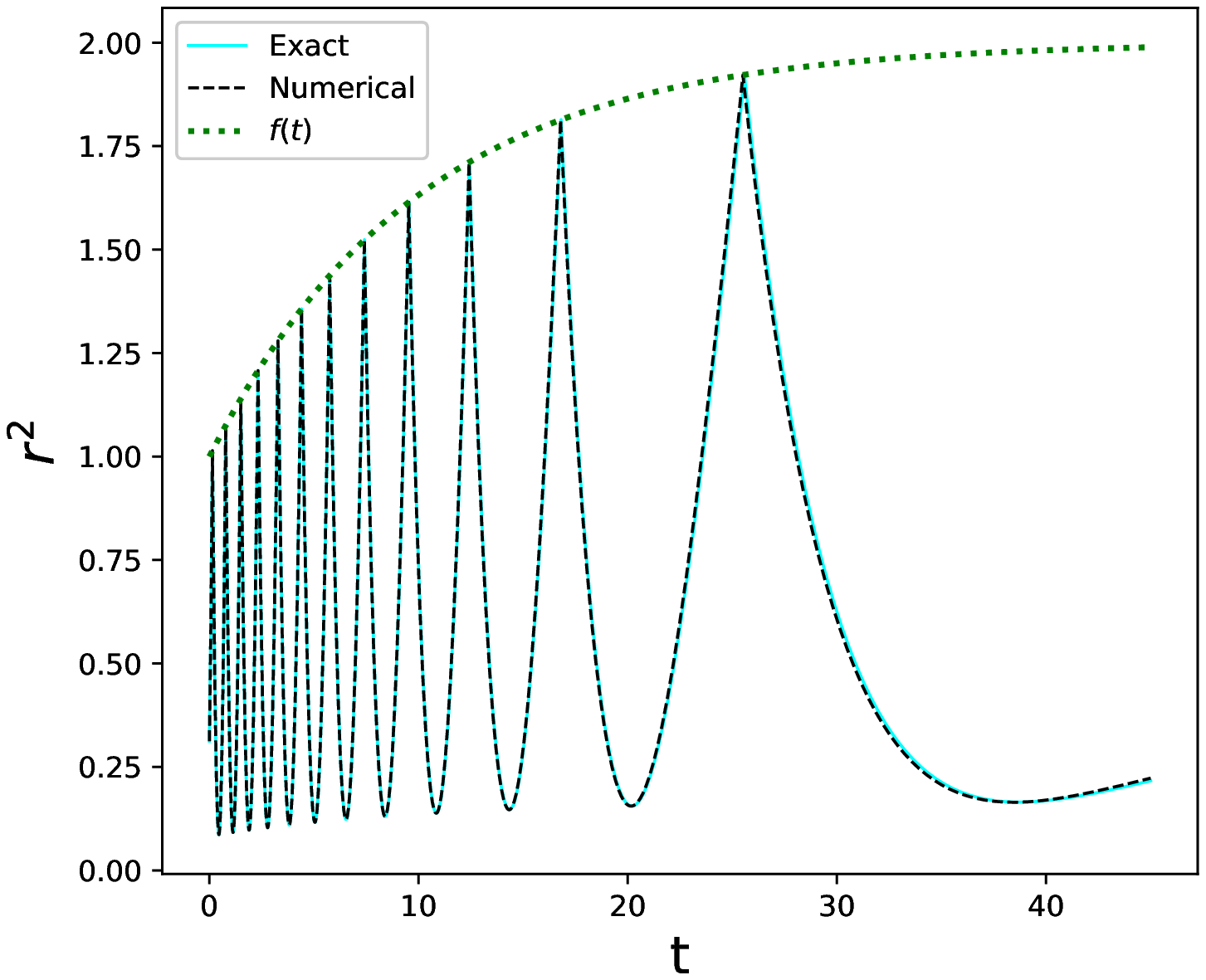}
	\centering\includegraphics[height=5.7cm]{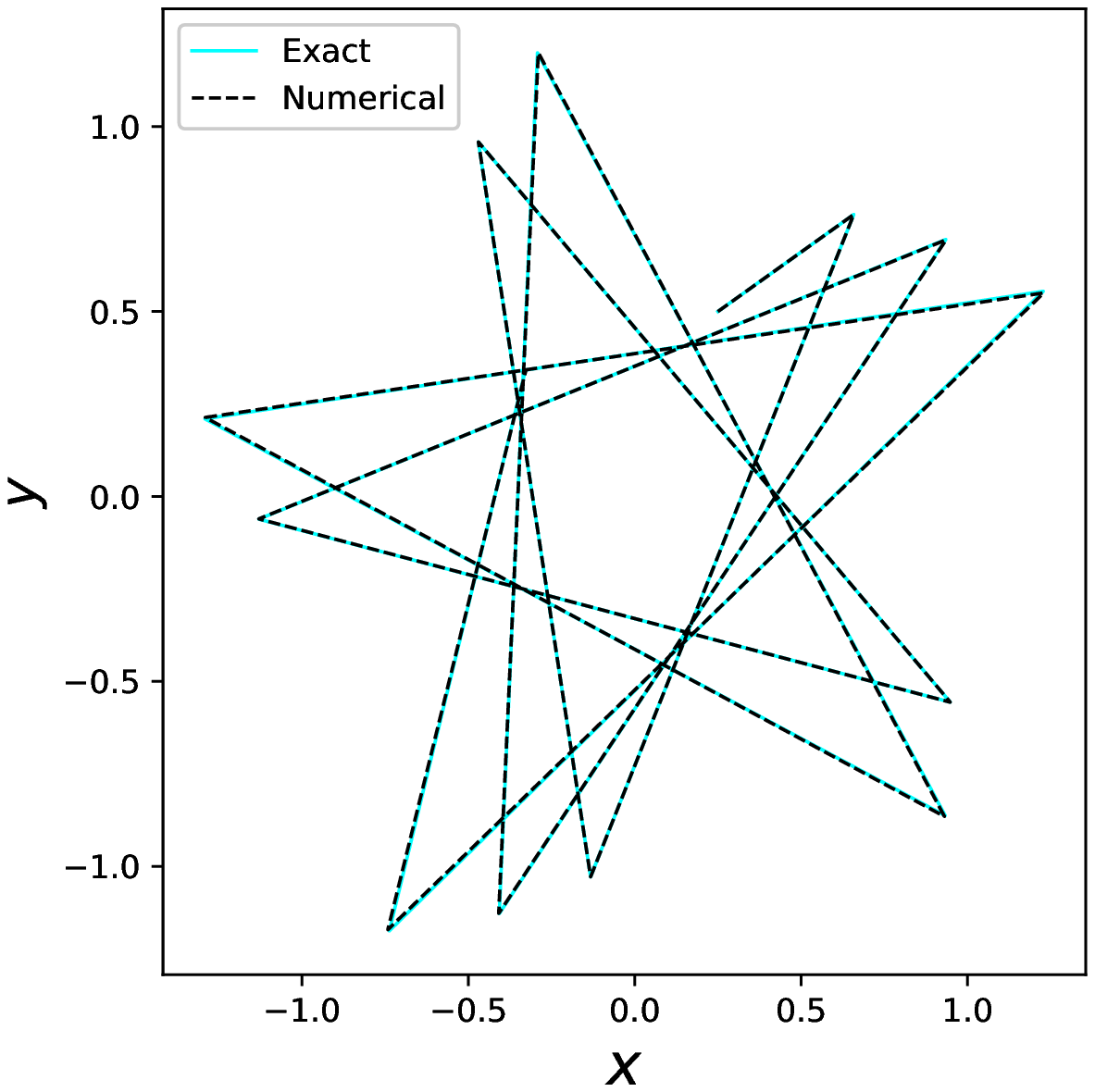}
	\caption{Simulation for $c=0.10$. he figure in the left corresponds with the reduced trajectory while the figure to the right corresponds with the reconstructed sulution}\label{fig:3}
\end{figure}

Using~\eqref{eq:Mom2}, the Routhian takes the form
\[
L_\mu(t,r,\dot{r})=\frac{1}{2}\exp{\left(\frac{c}{m}t\right)}m\dot r^2-\frac{\mu^2}{2mr^2}\exp{\left(-\frac{c}{m}t\right)}.
\]
The reduced reset map is determined by the expression~\eqref{eq:resetpolar} for $\dot r^+$ (note that the expression drops to the quotient since it only involves $r$, $\dot r$ and $f(t)$). The reduced switching surface is $S_\mu=\{r^2=f(t), \dot r>0\}$. One then obtains the following simple hybrid time-dependent Lagrangian system $\mathscr{L}=(Q_{\rm red},L_\mu,S_\mu,R_\mu)$, with $Q_{\rm red}\simeq \mathbb{R}^+$ parametrized by the radial coordinate $r$.
\begin{figure}[h]
	\centering\includegraphics[height=5.7cm]{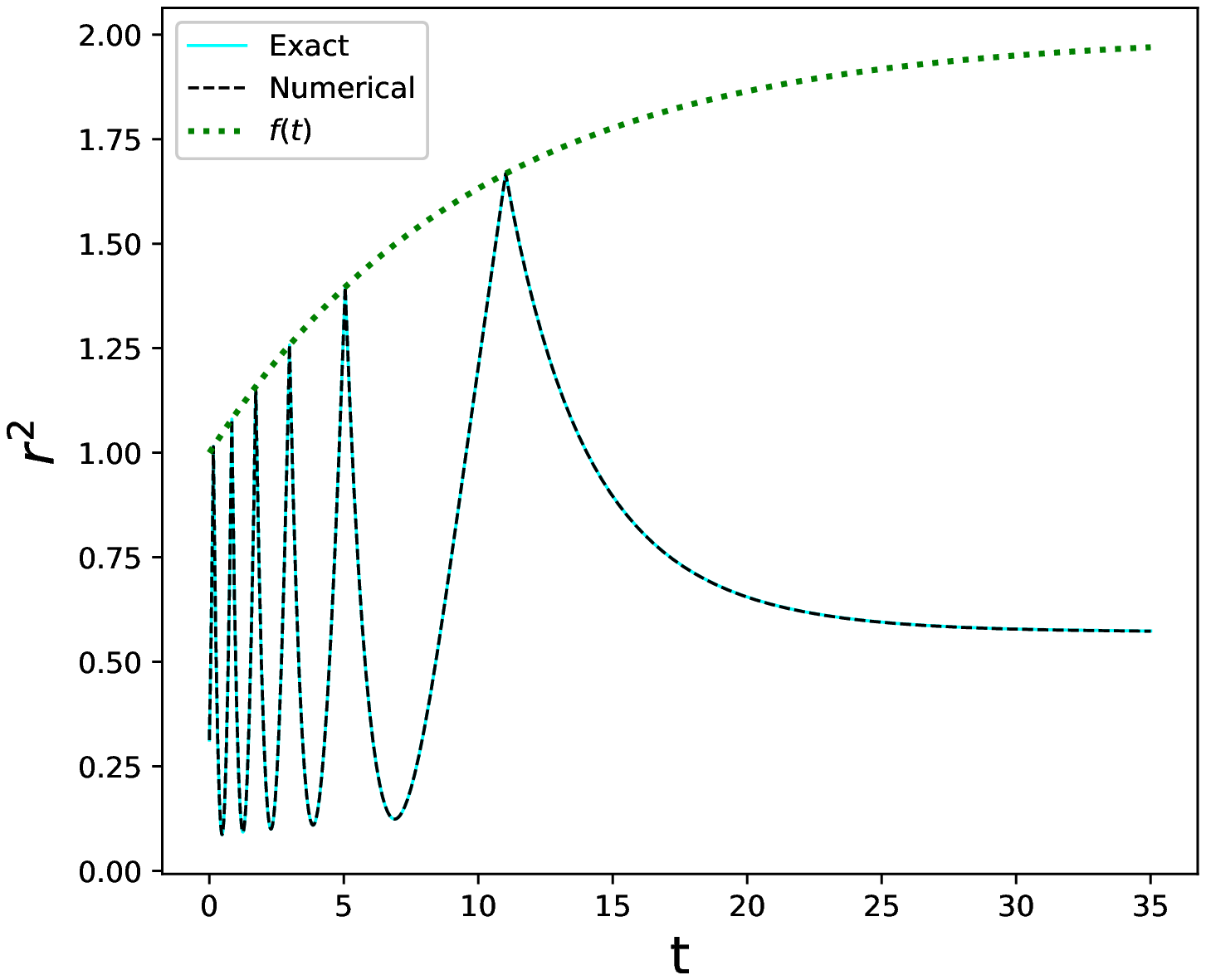}
	\centering\includegraphics[height=5.7cm]{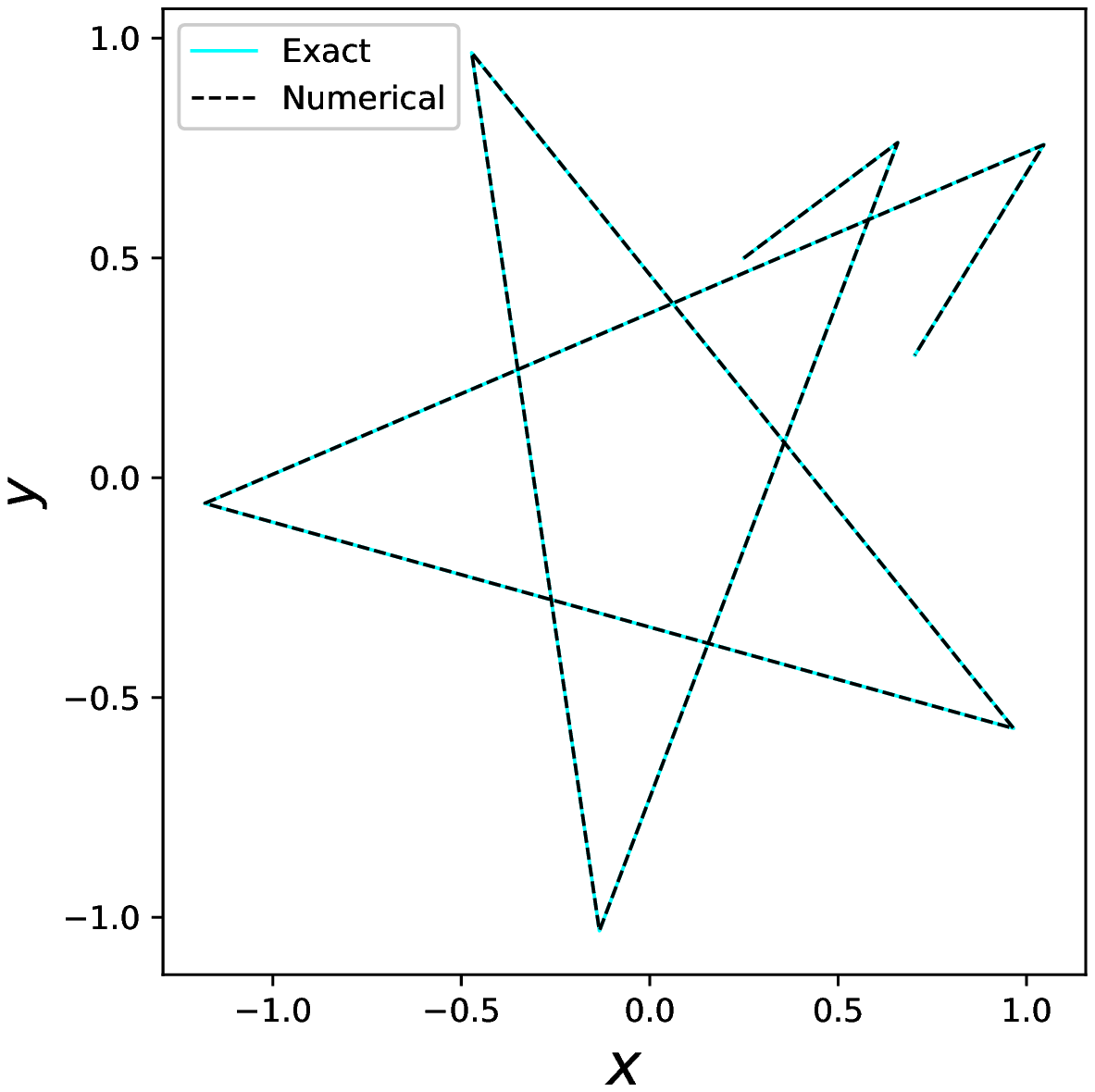}
	\caption{Simulation for $c=0.25$. The figure in the left corresponds with the reduced trajectory while the figure to the right corresponds with the reconstructed sulution}\label{fig:2}
\end{figure}
Figures~\ref{fig:2} and~\ref{fig:3} show numerical results using \textsc{Python} for two different values of the dissipation parameter $c$. The remaining parameters are the same for both simulations: $m=1$, $r(0)=0.5590$, $\dot r(0)=2.8621$, $\theta(0)=1.1071$ (rad), $\dot \theta(0)=-3.0400$ (rad/s), and the function $f(t)$ equals
\[
f(t)=2-\exp(t/10). 
\]
The reduced dynamics corresponding to $L_\mu$ is solved numerically (dashed black line) and used to integrate (numerically) the reconstruction equation~\eqref{eq:Mom2}
\[
\dot\theta=\exp{\left(-\frac{c}{m}t\right)} \frac{\mu} {m r^2}, 
\]
with $\mu$ determined from the initial conditions. Since~\eqref{eq:ex1} admits the explicit solution
\[
x(t)=x_0\exp{\left(-\frac{c}{m}t\right)},\quad y(t)=y_0\exp{\left(-\frac{c}{m}t\right)}, 
\]
a comparison with the analytical solution (solid teal line) can be easily obtained. Note, however,  that the impact times on which the particle bounces are obtained numerically also in this ``analytical'' case. 

\section{Conclusion}

We have extended Routh reduction to hybrid time-dependent mechanical systems. This extension recovers previous results on Routh reduction for hybrid systems in an intrinsic way, and hence broadens the applicability of the reduction procedure. The technique has been  illustrated both analitically and numerically in an illustrative example.

\section*{Appendix: Cosymplectic reduction theorem}
An \emph{almost cosymplectic structure} on a manifold $Q$ of odd
dimension $2n+1$ is a pair $(\eta,\Omega)$, where $\eta$ is a
$1$-form and $\Omega$ is a  $2$-form such that $\eta\wedge\Omega^n$
is a volume form on $Q$. The structure is said to be \emph{cosymplectic} if $\eta$ and $\Omega$ are closed. A manifold endowed with a cosymplectic structure is referred to as a cosymplectic manifold. 

A cosymplectic structure $(\eta,\Omega)$ on $Q$ induces
an isomorphism of $C^{\infty}(Q)$-modules
$\flat_{(\eta,\omega)}:\mathfrak{X}(Q)\to \Omega^1(Q)$ defined by
\begin{align*}\label{bemolle}
\flat_{(\eta,\Omega)}(X)=i_X\Omega+\eta(X)\eta,
\end{align*}
for every vector field $X\in \mathfrak{X}(Q)$. The vector field
$\mathbf{T}=\flat^{-1}_{(\eta,\Omega)}(\eta)$ on $Q$ is called the
\emph{Reeb vector field} of $(Q,\eta,\Omega)$ and it can
characterized by the following conditions
\[
i_\mathbf{T}\Omega=0,\qquad \eta(\mathbf{T})=1.
\] 

Let $G$ be a Lie group. An action $\phi:G\times Q\to Q$ of a Lie group $G$ on a cosymplectic
manifold $(Q,\eta,\Omega)$ is said to be \emph{cosymplectic} if
$\phi_g:Q\to Q$ is a cosymplectic map for any $g\in G$, i.e. if 
\[
\phi_g^*\eta=\eta,\quad  \phi_g^*\Omega=\Omega.
\]
Given a cosymplectic action on $Q$, a smooth map $J:Q\to\mathfrak{g}^*$ is said to be a \emph{momentum map} if
the infinitesimal generator $\xi_Q\in\mathfrak{X}(Q)$ of the action associated
with any $\xi\in\mathfrak{g}$ is the Hamiltonian vector field of the function
$J_\xi:Q\to\mathbb{R}$ defined by the natural pointwise pairing $J_\xi(q)=\langle J(q),\xi \rangle$. The momentum map is \emph{$\hbox{Ad}^*$-equivariant} (or \emph{equivariant}, for short) if it is equivariant
with respect to the action $\phi$ and to the coadjoint action
$\hbox{Ad}^*:G\times\mathfrak{g}^*\to\mathfrak{g}^*$, that is
\[
J(\phi_g(q))=\hbox{Ad}^*_{g^{-1}}(J(q)).
\]

Let $\mu\in\mathfrak{g}^*$ be a weakly regular value of an equivariant
momentum map $J:Q\to\mathfrak{g}^*$. We denote by $G_\mu$ the isotropy group of $\mu$ with respect to the
coadjoint action, i.e. $G_\mu=\{g\in G:\,\hbox{Ad}^*_{g}\mu=\mu\}$.

The action $\phi$ induces an action $\phi:G_\mu\times J^{-1}(\mu)\to J^{-1}(\mu)$ of $G_\mu$ on the submanifold $J^{-1}(\mu)$. Following \cite{Albert} we will say that this action is
quotientant if the
orbit space $Q_\mu=J^{-1}(\mu)/G_\mu$ admits a smooth manifold
structure and the canonical projection $\pi_\mu:J^{-1}(\mu)\to
Q_\mu$ is a surjective submersion.

\begin{theorem*}[Cosymplectic reduction Theorem, \cite{Albert}]\label{CosymplecticReduction}
	Let $\phi:G\times Q\to Q$ be a  cosymplectic action of a Lie group
	$G$ on a cosymplectic manifold $(Q, \eta,\Omega)$. Suppose that
	$J:Q\to\mathfrak{g}^*$ is an $\hbox{Ad}^*$-equivariant momentum map associated with
	$\phi$ such that $\mathbf{T}(J)=0$ where $\mathbf{T}$ is the Reeb vector field of $Q$. Let $\mu\in\mathfrak{g}^*$ be a weakly 
	regular value of $J$ such that the induced action of $G_\mu$
	on $J^{-1}(\mu)$ is quotientant. Then, $Q_\mu=J^{-1}(\mu)/G_\mu$ is a
	cosymplectic manifold with cosymplectic structure
	$(\eta_\mu,\Omega_\mu)$ characterized by
	\begin{equation*}\label{ReducedCosymplecticStruct}
	\pi_\mu^*\eta_\mu=i_\mu^*\eta, \qquad
	\pi_\mu^*\Omega_\mu=i_\mu^*\Omega,
	\end{equation*}
	where $\pi_\mu:J^{-1}(\mu)\to Q_\mu$ is the canonical projection and
	$i_\mu:J^{-1}(\mu)\hookrightarrow Q$ is the canonical inclusion.
	
	Moreover, the restriction $\mathbf{T}|_{J^{-1}(\mu)}$ of $\mathbf{T}$ is tangent
	to $J^{-1}(\mu)$ and $\pi_\mu$-projectable onto the Reeb vector field
	$\mathbf{T}_\mu$ of $Q_\mu$.
\end{theorem*}

As a matter of fact, the cosymplectic reduction theorem of Albert  \cite{Albert}  can be obtained directly from the Marsden-Weinstein reduction theorem \cite{MaWe}, see \cite{singular}.

\section*{Acknowledgments}

The authors are indebted to Andrea Bel for the assistance with \textsc{Python}, and to David Mart\'in de Diego for the fruitful discussions. EGTA and EEI are grateful to the ICMAT (Madrid) for its hospitality during the visits which made this work possible.

\end{document}